\documentclass[%
showpacs,
 amsmath,amssymb,
 aps,
prl,
twocolumn,
floatfix,
]{revtex4-1}

\usepackage[normalem]{ulem}
\usepackage{graphicx}
\usepackage{dcolumn}
\usepackage{epstopdf}
\usepackage{bm}
\usepackage{enumerate}
\usepackage{fancyhdr}
\usepackage[usenames, dvipsnames]{color}
\usepackage{tikz}
\usepackage{ragged2e}
\usetikzlibrary{positioning}
\usetikzlibrary{arrows, decorations.markings}
\usetikzlibrary{decorations.pathmorphing,shadows} 
\usepackage{bbm}

\usepackage[utf8]{inputenc}
\usepackage{amssymb}
\usepackage{amsmath}
\usepackage{mathtools,etoolbox}
\usepackage{pgf,tikz}\usepackage{mathrsfs}\usetikzlibrary{arrows}
\usepackage{nonfloat}
\usepackage{mwe}
\usepackage{upgreek}	

\begin{document}

\preprint{APS/123-QED}

\title{Two-particle sub-wavelength Quantum Correlation Microscopy}

\author{Josef G. Worboys, Daniel W. Drumm, and Andrew D. Greentree}
 \affiliation{Australian Research Council Centre of Excellence For Nanoscale Biophotonics, RMIT University, Melbourne 3001, Australia.}
\email{josefgworboys@gmail.com}
\email{\mbox{Andrew.Greentree@rmit.edu.au}}





\date{\today}

\begin{abstract}
Typically, optical microscopy uses the wavelike properties of light to image a scene. However, photon arrival times provide more information about emitter properties than the classical intensity alone. Here, we show that the Hanbury Brown and Twiss experiment (second-order correlation function) measures the intensity asymmetry of two single photon emitters, and that by combining the total number of detected photons with the zero-lag value of the correlation function, the positions and relative brightness of two emitters in two dimensions can be resolved from only three measurement positions -- trilateration, a result that is impossible to achieve on the basis of intensity measurements alone. 
\end{abstract}


\maketitle

\textit{Introduction.---}Optical microscopy is one of the most important tools for the understanding of biological materials.  Conventional microscopy is limited by the wavelike nature of light, and the associated diffraction limit which imposes a fundamental restriction on imaging resolution in the absence of \textit{a priori} information of the system being imaged.  There are now many techniques to beat the diffraction limit,  defining the emerging field of optical nanoscopy.  Such techniques either use nonlinear optical processes, blinking, \textit{ans\"{a}tzer} about the system, or quantum techniques to derive subwavelength information.  Interestingly, all of the superresolution techniques \cite{Hell:94,Rust2006,Sauer3505} appear to have qualitatively similar scaling laws of resolution with probe intensity, suggesting a common framework exists for understanding the fundamental limits of resolution in microscopy \cite{Hemmer2016,RN46}.

Quantum correlations are an intriguing resource for optical nanoscopy.  The Hanbury Brown and Twiss (HBT) experiment was initially developed as a method of determining stellar parallax \cite{Brown1956} and was the first experiment to definitively prove the quantisation of the electromagnetic field.  In its simplest form, HBT uses two single-photon detectors that receive signals from the same site.  When an emitter produces no more than one photon at any given time, quantisation of the field means that the photon cannot be detected at both detectors simultaneously, and therefore the cross correlation signal must go to zero at zero time delay. 

Quantum correlation microscopy uses measurements derived from HBT signals across multiple emitters to improve microscopy resolution.  This scheme was first introduced by Schwarz and Oron via a wide-field solution \cite{SchwartzOrenLevitt} and demonstrated by Monticone \textit{et al.} using nitrogen-vacancy colour centres in diamond \cite{Monticone} with confocal microscopy. In its simplest form, the correlation function provides information about the number of emitters, which can be used in centroid-type fitting algorithms \cite{Cheezum01,Mello16,Yoshihiro}.
These works show that using quantum correlations including correlations higher than two, provides an improvement in confocal and widefield resolution that scales as $\sqrt{k}$ where $k$ is the order of the quantum correlation. 

Trilateration is the determination of the position of an object on the basis of the intersection of structures with circular symmetry and finds practical application in fields such as surveying and satellite global positioning systems \cite{Baun,Joon}. The task of locating a single photon emitter or emitters on a two-dimensional plane using confocal microscopy \cite{RobertWebb,VanKempen,Jonkman2015} is also a problem in lateration \cite{Hightower01locationsensing}, although it is usually not investigated as such. Confocal microscopy is lateration in that a circularly symmetric point-spread function (PSF) is scanned across a scene, and the emitted light is collected as a single intensity value through the same point-spread function.  For brevity,  we use PSF to represent the product of the illumination and collection point-spread functions. The reconstruction of an image based on the circularly symmetric PSFs is properly the lateration step.  Understanding the minimum requirements on the number of measurement locations (PSFs)  is important as conventional scanning confocal techiques are known to be suboptimal for search \cite{DrummGreentree2017}.  

Here we show that quantum measurements from three locations suffice to determine the location of two particles separated by less than the diffraction limit, in two dimensions.  Such a determination on the basis of intensity-only measurements is impossible as there are five unknown quantities (the $x$ and $y$ locations of each emitter and the relative brightness of the emitters) and only three measurement results plus the constraint of only two emitters in the field of view.  We construct explicit simulations of the HBT cross-correlation function for the case of two emitters of unequal brightness, and show the origin of the increased information relative to simple measurements of the intensity.  We focus on the ideal case here; however, this work is specifically useful in studying problems such as dimerisation, which is important in certain biochemical reactions \cite{BodnarUrszulaPoleszczuk,WebbHirschNeedham,DominguesHirschTynan,DonlanCosterton}.

This paper is organised as follows.  We begin by describing the HBT experiment for two emitters, highlighting the cross-correlation function as measure of the brightness asymmetry of the emitters. We then show the trilateration problem for two emitters including the achievable resolution under practical considerations.

\textit{Cross-correlation function for two particles.}---%
The HBT experiment measures photon detection coincidences between two detectors monitoring the same spatial region as a function of time.  

The normalised coincidence rate is expressed as a function $g^{(2)}$ of the time delay $\uptau$ between photon detections at the different detectors, where the number of coincidences detected at $\uptau$ is normalised by the number of uncorrelated coincidences at $ \uptau = \pm \infty$ which is a measure for the total number of single photon clicks; $g^{(1)}$. For our purposes we are most concerned with the coincidences at $\uptau = 0$, \textit{i.e.}, $g^{(2)}(0)$.

\begin{figure}[tb!]
\includegraphics[width=0.8\columnwidth]{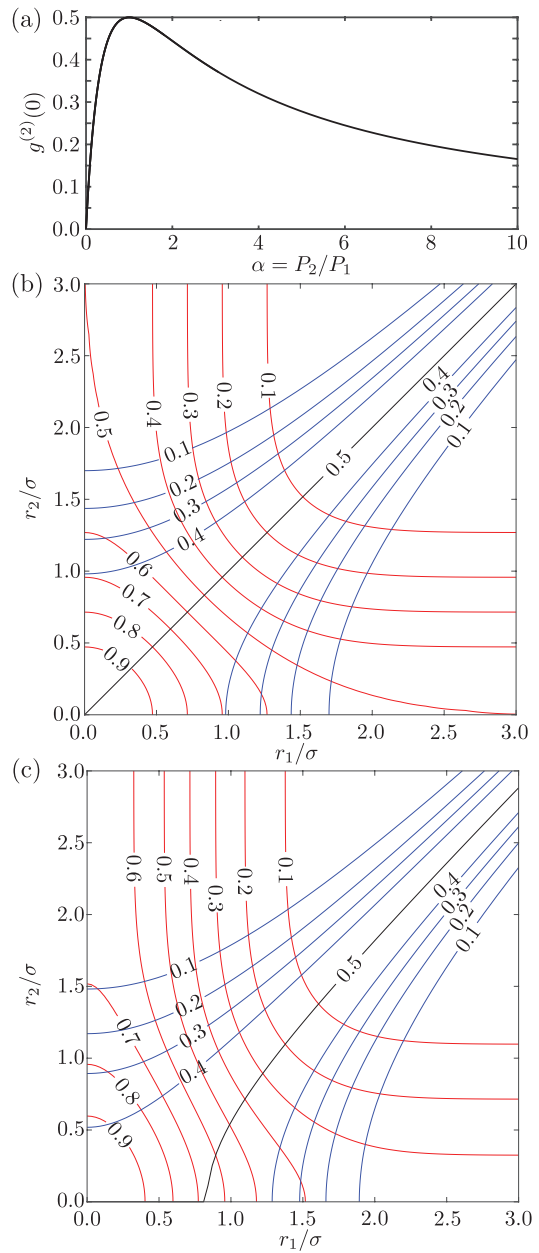}
\caption{(a) Cross correlation function $g^{(2)}(0)$ for two particles as a function of relative brightness, $\alpha$. The maximum value of $g^{(2)}(0) = 0.5$ is achieved for equal brightness particles. (b) Overlapped contour plots of $g^{(1)}(r_1,r_2,\alpha)$ ({\color{red} red}) and $g^{(2)}(\uptau=0,r_1,r_2,\alpha)$ ({\color{blue} blue}) for $\alpha=1$, and (c) $\alpha=0.5$ where radial distance is measured in units of the standard deviation of the point spread function of the illumination/collection optics.  By comparing both the $g^{(1)}$ and $g^{(2)}$ values, it is possible to determine more information about the particles' positions than is possible using intensity alone.
 When $\alpha\neq 1$ the symmetry between the contours for $r_1$ and $r_2$ is broken. 
}
\label{fig:g2vsalpha}

\end{figure}

It is often erroneously claimed that $g^{(2)}(0)< 0.5$ implies that one single photon emitter is being observed. This is due to the well-known result that for $n$ co-located emitters of equal brightness $g^{(2)}_n(0) = (1-1/n)$.  
However this result does not hold in the case that the intensities measured from the emitters are not equal, nor does it hold for equally bright, spatially-separated emitters.  For two emitters of unequal brightness we find 
\begin{align}
g^{(2)}_2(0) = \frac{2P_1 P_2}{\left(P_1 + P_2\right)^2} = \frac{2\alpha}{\left(1 + \alpha\right)^2}, \label{eq:g2arb}
\end{align}
where we have introduced $\alpha$ = $P_2/P_1$   as the ratio of the probability $P_1$ of detecting a photon from particle 1, divided by the probability $P_2$ of detecting a photon from particle 2.  As the probability of photon detection is directly proportional to the received power from a given emitter, this result demonstrates the role of the Hanbury Brown-Twiss measurement in determining  brightness asymmetry.  The difference in brightness can be from any cause, for example because the two emitters are of different species.  Equally, however, the difference in brightness could be due to the two emitters being located at different positions relative to the centre of the PSF used to interrogate them, and it is this that provides a novel method of localisation. A plot showing the relation between $\alpha$ and $g^{(2)}_2(0)$ is shown in Fig.~\ref{fig:g2vsalpha}(a), which highlights the singular case where $g^{(2)}_2(0)$ achieves its maximum value of 0.5: when $\alpha = 1$. 

We define $P_{0i}$ as the maximum probability of detecting a photon from emitter $i = 1,2$, when that emitter is positioned at the centre of the detection point spread function.  $P_{0i}$ is directly proportional to the brightness of the emitter, and we assume that the overall measurement efficiency of the microscope is the same for each particle.  For simplicity we assume that the microscope PSF can be treated as a Gaussian, which is known to be good for two dimensions \cite{Zhang:07}.  Although such a treatment is not ideal for practical microscopy\cite{Stallinga:10}, it serves to illustrate our method, and the use of more complicated PSFs will not alter our results significantly.

The probability of detecting a photon from emitter $i$ is
\begin{align}
P_i = \left(P_{0i}/\sqrt{2\pi\sigma^2}\right) \exp\left[-r_i^2/\left(2\sigma^2\right)\right], \label{eq:prob}
\end{align}
where $r_i$ is the distance from the emitter to the origin of the detection point spread function and $\sigma \approx 0.21 \lambda/NA$ is the standard deviation of the effective Gaussian point spread function for wavelength $\lambda$ and numerical aperture $NA$. 
By comparing $g^{(2)}_2(0)$ with the total  intensity, or sum of probabilities of detecting photons at a given time, $g^{(1)} = P_1 + P_2$ 
, we observe that the two techniques provide qualitatively different information. Figure \ref{fig:g2vsalpha} (b) shows $g^{(1)}$ with $g^{(2)}_2(0)$ as functions of $r_1$ and $r_2$ for the case $\alpha = 1$. A given measurement of both $g^{(1)}$ and $g^{(2)}_2(0)$ therefore determines both $r_1$ and $r_2$ up to ambiguity in the labeling.  The case is only slightly more complicated for $\alpha \neq 1$, [Fig.~\ref{fig:g2vsalpha} (c)], which shows the corresponding contours for $\alpha = 0.5$, where the symmetry between $r_1$ and $r_2$ is broken, however there are still two valid $(r_1,r_2)$ pairs that will satisfy the $g^{(1)}$ and $g^{(2)}_2(0)$ data.

\textit{Quantum trilateration.}---%
We now proceed to invert the analysis in the previous section to determine the location of the two emitters on the basis of three Hanbury Brown and Twiss measurements: quantum trilateration. 

On the basis of three measurement locations and six measurement outcomes [three intensity and three $g_2^{(2)}(0)$] taken at the detector locations, our approach computes the least square error between predicted values of the measured quantities and trial locations for two emitters. 
We stress that the intensity result can be obtained simply from the square root of the number of coincidences obtained at $ \uptau = \pm \infty$, 
and so does not require an additional detector or detector channel.  Alternatively, all of the required data could be obtained from a single-photon-resolving detector at each measurement location where $g^{(2)}_2(0)$ is given by the number of two photon measurements per unit time divided by the square of the number of single photon measurements per unit time.  Nevertheless, the standard HBT two-detector setup is the most common available experimental apparatus for performing such measurements, so that is our focus here.

We do not attempt to optimise the location of the detectors for our purpose, we simply note the following heuristics for their placement relative to the two emitters.  For our scheme to be beneficial compared to standard methods, we require the two emitters to be close with respect to the diffraction limit, so as to be unresolved by conventional means, but so that we obtain a significant number of coincidence detections.  For simplicity, we assume that the detector locations are placed at the vertices of an equilateral triangle spaced one standard deviation $\sigma$ apart based on the illumination PSF.

Our simulation stochastically assigns real space locations to two single-photon emitters inside one standard deviation of the PSF, $\sigma$, and all of our position results are therefore scaled in units of $\sigma$.  Additionally, we randomly choose the relative power of the emitters, $0\leq \alpha \leq 1$, without loss of generality ($\alpha > 1$ are equivalent under an emitter label swap). To take into account realistic experimental noise, we assume that the noise in determining $g^{(2)}(0)$ is based on counting noise, and therefore express the uncertainty in the obtained data (total counts or coincidences) as $\eta = 1/\sqrt{N}$ where $N$ is the number of detected coincidences for $g^{(2)}(0)$.  For simplicity we also use this same value to apply noise to $g^{(1)}$; we note this is a considerable overestimation of the intensity noise.

Using the analytical results from Eqs.~\ref{eq:g2arb} and \ref{eq:prob}, our code adds relative errors at the level $\eta$, i.e. for each detector position, $j=1,2,3$ we generate
\begin{align}
G_j^{(i)} = \left[1+\eta \texttt{randn}\right]g_j^{(i)} \label{eq:randn}
\end{align}
where $\texttt{randn}$ is the \textsc{Matlab} \cite{MATLAB:2018} function generating a normally distributed random value with mean 0 and standard deviation, $G_{j}^{(i)}$ are the simulated measurement results and we have dropped the $(0)$ from the $g^{(2)}(0)$ and $G^{(2)}(0)$ for simplicity. On the basis of these synthetic measurement results, we then attempt to determine the emitter positions and relative intensity that minimises the sum of the squared errors, i.e. 

\begin{align}
\chi^2 = \Sigma_{i,j}\left[\mathcal{G}_j^{(i)}(x_1,y_1,x_2,y_2,\alpha) - G_j^{(i)}\right]^2, \label{eq:ChiSquared}
\end{align}
where
$\mathcal{G}_j^{(i)}(x_1,y_1,x_2,y_2,\alpha)$ is the expected value of $i=1$ intensity or $i=2$ HBT coincidence, for detector $j$ for the trial values $x_1...\alpha$.  

\begin{figure}[b!] 
\includegraphics[width=0.8\columnwidth]{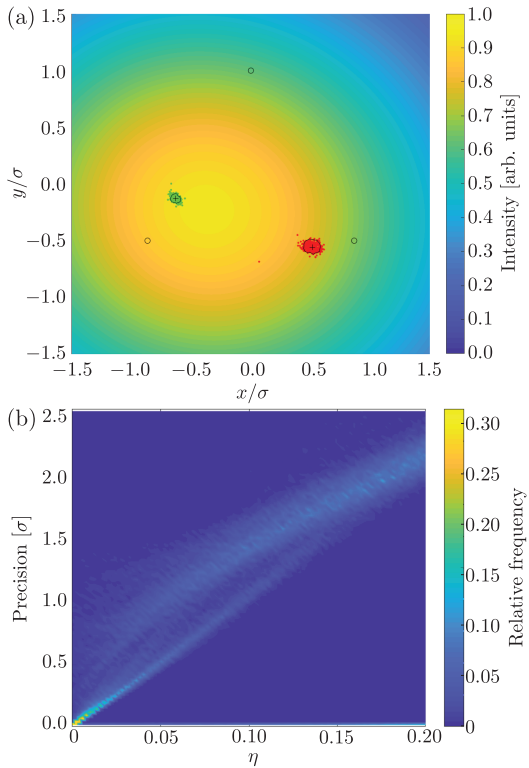}
\caption{(a) Results of trilateration for two single-photon emitter locations $(x_1,y_1) = (-0.6300, -0.1276)$, $(x_2,y_2) = (0.5146, -0.5573)$ ratio of emitter power ($\alpha = 0.3617$), and system noise/error $\eta = 0.01$. The pseudo-color plot shows the predicted confocal map for the case of a single detector scanned across the entire region, the black open circles the detector locations for trilateration, the crosses are the emitter locations, and the green and red dots are the calculated locations of emitter 1 and 2 respectively, with $90\%$ of the computed positions shown by the black contours around the emitter locations. (b) Histograms showing the precision achieved after applying the trilateration protocol for randomly chosen emitters at each noise value $\eta$.  The colour of each precision-$\eta$ point shows the proportion of times that precision was achieved under those noise conditions.  Noticeable is the presence of two bands within the results.  The lower straight line is obtained for $\alpha \lessapprox 0.5$, and the upper band is obtained for cases where $0.5 \lessapprox \alpha \lessapprox 1$.  Cases with $\alpha\lessapprox 0.05$ have been excluded as discussed in the text.
} 
\label{fig:TrilaterationFig2}
\end{figure}

The results of performing such a trial are shown in Fig.~\ref{fig:TrilaterationFig2}(a) where two emitters were randomly placed at $(x_1,y_1) = (-0.6300, -0.1276)$, $(x_2,y_2) = (0.5146, -0.5573)$ ratio of emitter power ($\alpha = 0.3617$), and system noise/error $\eta = 0.01$.  The figure shows  in color the expected results of performing a standard confocal map across the scene (normalised to the maximum power of emitter 1, $P_{0,1}$.  As can be seen, the two emitters are not resolved.

For the purposes of quantum trilateration, we consider three detector positions, $(0,1)$, $(\sqrt{2},-0.5)$ and $(-\sqrt{2},-0.5)$.  We then computed 501 trials using the method outlined in Eq.~\ref{eq:randn} and independently determined the expected emitter locations using the \textsc{Matlab} routine \texttt{fminsearch} to gauge for the error.  The results of each individual run are shown by the green and red dots for emitter 1 and emitter 2 respectively, which accord well with the true locations.  Due to statistical effects, it is always expected that there will be some error in retrieving the true locations, and occasional, pathological 
cases can greatly skew the fitting \cite{Hlupic2013ADA}.  Accordingly for each emitter, we have shown the boundary 
that contains the $90\%$ of the 501 fitted data points that are closest to the determined mean location, and we define the precision for each emitter's location as the maximum radius of this boundary.

To quantify the scaling of our protocol's precision with respect to detection noise, we show a histogram (Fig. \ref{fig:TrilaterationFig2}b) of the summed precision across both emitters for an ensemble of 639 randomly chosen emitter locations and relative brightnesses $(x_1,y_1,x_2,y_2,\alpha)$, trialled 501 times apiece per value of system noise $0\leq\eta\leq 0.20$.  When $\alpha \lessapprox 0.05$ our protocol fails to locate the less bright emitter, although the brighter emitter is localised.  Accordingly we have removed these unconverged cases from our dataset, leaving 639 of the 665 cases feasible under supercomputer walltime limits.

Clearly noticeable are two distinct bands of data.  
The lower of the two bands corresponds to the case where $\alpha \lessapprox 0.5$, which demonstrates a linear scaling 
of localisation with $\eta$ or equivalently with $1/\sqrt{N}$ as the main source of uncertainty with $\eta$ is counting statistics.  The upper band, attained for $0.5 \lessapprox \alpha < 1$, is more interesting.  We see that for comparable $\eta$ the localisation error is greater.  This indicates that when there is a large difference in the intrinsic brightness of the emitters, the protocol shows better scaling than when the two emitters are of comparable brightness.  

\textit{Conclusions.}---%
Our results show that by combining quantum correlation with conventional intensity measurements, it is possible to solve the quantum trilateration problem for two particles of unknown relative intensity: a problem impossible to solve on the basis of intensity measurements alone.  This methodology highlights the critical information accessible to Hanbury Brown and Twiss measurements that is not present in conventional confocal-type measurements, thereby clarifying the origin of the speedup seen in quantum microscopy \cite{Hell:94,Rust2006,Sauer3505,SchwartzOrenLevitt,RN19,Muthukrishnan2004,IsraelOronSilberberg}.

Furthermore, by showing that only three measurement locations are required for superresolution localisation of two particles, our results are significant in the search for optimal strategies for microscopy.  Optimal microscopy is necessary as effects such as phototoxicity limit the application of superresolution methods in biology, and it is therefore necessary to quantify the total photon budget necessary to obtain a desired resolution in any experiment.  

We note that our results have been normalised in units of the point spread function of the excitation field.  However, although we have compared our results with standard confocal microscopy, there is in fact no restriction on the microscopy technique.  So for example, the quantum trilateration approach could be combined with other superresolution techniques, for example STED microscopy, and our approach would provide commensurate increases to the obtained STED resolution as shown above.  

Lastly, we note that we have not attempted to optimise the locations of the detectors relative to the emitters, nor have we considered the advantage of increasing the number of detectors beyond three.  We leave this optimisation for future work.

The authors acknowledge useful conversations with Brant Gibson, Antony Orth and Ewa Goldys, as well as the support of the ARC Centre of Excellence for Nanoscale BioPhotonics (CNBP) (Grant No.CE140100003), JGW acknowledges RMIT \& CNBP for Ph.D scholarship funding.  ADG acknowledges the support of an ARC Future Fellowship (Grant No. FT160100357). This research was undertaken with the assistance of resources and services from the National Computational Infrastructure (NCI), which is supported by the Australian Government (LE160100051).

\bibliographystyle{apsrev4-1}

%

\end{document}